\documentclass[conference]{IEEEtran}
\IEEEoverridecommandlockouts
\pdfoutput=1
\usepackage{graphicx}
\usepackage{amssymb}
\usepackage{mathtools}
\usepackage{dsfont}
\usepackage{cite}
\usepackage{stfloats}
\usepackage{subfigure}
\usepackage{psfrag}
\usepackage[mathscr]{euscript}
\usepackage{acronym}  % make an acronym
\usepackage{amsmath}
\usepackage{algorithm}
\usepackage[noend]{algpseudocode}
\usepackage{booktabs}
\usepackage{bbm}

\makeatletter
\def\BState{\State\hskip-\ALG@thistlm}
\makeatother

\usepackage{float}
\usepackage{balance}

\acrodef{CCDF}{complementary cumulative distribution function}
\acrodef{CF}{characteristic function}
\acrodef{PPP}{Poisson point processe}
\acrodef{RV}{random variable}
\acrodef{i.i.d.}{independent and identically distributed}
\acrodef{PDF}{probability distribution function}
\acrodef{CDF}{cumulative distribution function}
\acrodef{ch.f.}{characteristic function}
\acrodef{AWGN}{additive white Gaussian noise}
\acrodef{SNR}{signal-to-noise ratio}
\acrodef{LRT}{likelihood ratio test}
\acrodef{DRT}{distance ratio test}
\acrodef{GLRT}{generalized likelihood ratio test}
\acrodef{CRLB}{Cram\'{e}r-Rao lower bound}
\acrodef{CRB}{Cram\'{e}r-Rao bound}
\acrodef{ZZLB}{Ziv-Zakai lower bound}
\acrodef{ZZB}{Ziv-Zakai bound}
\acrodef{LOS}{line-of-sight}
\acrodef{ToF}{time-of-flight}
\acrodef{NLOS}{non-line-of-sight}
\acrodef{GDOP}{geometric dilution of precision}
\acrodef{GPS}{Global Positioning System}
\acrodef{FIM}{Fisher information matrix}
\acrodef{PEB}{position error bound}
\acrodef{SPEB}{squared position error bound}
\acrodef{TOA}{time-of-arrival}
\acrodef{TOF}{time-of-flight}
\acrodef{WSN}{wireless sensor network}
\acrodef{MAC}{medium access control}
\acrodef{RSS}{received signal strength}
\acrodef{WAF}{wall attenuation factor}
\acrodef{TDOA}{time difference-of-arrival}
\acrodef{RF}{radiofrequency}
\acrodef{RTT}{round-trip time}
\acrodef{AOA}{angle-of-arrival}
\acrodef{MF}{matched filter}
\acrodef{ED}{energy detector}
\acrodef{ML}{maximum likelihood}
\acrodef{MSE}{mean-square error}
\acrodef{RMSE}{root-mean-square error}
\acrodef{LEO}{localization error outage}
\acrodef{ppm}{part-per-million}
\acrodef{ACK}{acknowledge}
\acrodef{UWB}{Ultrawide bandwidth}
\acrodef{TNR}{threshold-to-noise ratio}
\acrodef{LS}{least squares}
\acrodef{IR-UWB}{impulse radio UWB}
\acrodef{FCC}{Federal Communications Commission}
\acrodef{TH}{time-hopping}
\acrodef{PPM}{pulse position modulation}
\acrodef{MUI}{multi-user interference}
\acrodef{PDP}{power delay profile}
\acrodef{BPZF}{band-pass zonal filter}
\acrodef{SIR}{signal-to-interference ratio}
\acrodef{SINR}{signal-to-interference-plus-noise ratio}
\acrodef{RFID}{radio frequency identification}
\acrodef{WPAN}{wireless personal area network}
\acrodef{WWB}{Weiss-Weinstein bound}
\acrodef{DP}{direct path}
\acrodef{MF}{matched filter}
\acrodef{MMSE}{minimum-mean-square-error}
\acrodef{SBS}{serial backward search}
\acrodef{SBSMC}{serial backward search for multiple clusters}
\acrodef{NBI}{narrowband interference}
\acrodef{WBI}{wideband interference}
\acrodef{INR}{interference-to-noise ratio}
\acrodef{CR}{channel response}
\acrodef{CIR}{channel impulse response}
\acrodef{CR}{channel  response}
\acrodef{RADAR}{radar}
\acrodef{MUR}{Multistatic radar}
\acrodef{JBSF}{jump back and search forward}
\acrodef{HDSA}{high-definition situation-aware}
\acrodef{RRC}{root raised cosine}
\acrodef{ST}{simple thresholding}
\acrodef{BTB}{Bellini-Tartara bound}
\acrodef{P-Max}{$P$-Max}  %suggestion, use with \acl{P-Max}
\acrodef{MIMO}{multiple-input multiple-output}
\acrodef{MAP}{maximum a posteriori}
\acrodef{FG}{factor graph}
\acrodef{OP}{outage probability}
\acrodef{WED}{wall extra delay}
\acrodef{RMS}{root mean square}
\acrodef{SPAWN}{sum-product algorithm over a wireless network}
\acrodef{MDD}{minimum distance distribution}
\acrodef{MAP}{maximum a posteriori probability}
\acrodef{SAP}{small cell access point}
\acrodef{UE}{user equipment}
\acrodef{MBS}{macro cell base station}
\acrodef{UER}{\ac{UE} Relay}
\acrodef{D2D}{device-to-device}
\acrodef{MBS}{macro base station}
\acrodef{CSI}{channel state information}
\acrodef{OGR}{outage guard region}
\acrodef{FUR}{feasible UER region}
\acrodef{EHR}{energy harvesting region}
\acrodef{EH}{energy harvesting}
\acrodef{D2D-EHSN}{D2D communication provided \ac{EH} small cell network}
\acrodef{D2D-EHHN}{D2D communication provided \ac{EH} heterogeneous network}
\acrodef{3GPP}{3rd Generation Partnership Project}
\acrodef{BS}{base station}
\acrodef{DF}{decode and forward}
\acrodef{CCDF}{complementary cumulative distribution function}
\acrodef{ZF}{zero forcing}
\acrodef{RZF}{regularized zero forcing}
\acrodef{WLLN}{weak law of large number}
\acrodef{SLLN}{strong law of large numbers}
\acrodef{TDD}{Time-division duplex}
\acrodef{EE}{energy efficiency} 
\acrodef{HetNet}{heterogeneous network} 
\acrodef{SCP}{Single Cell Processing}
\acrodef{CBF}{Coordinated Beamforming}
\usepackage{color}
\usepackage{dsfont}
\usepackage{bbm}

\DeclareMathAlphabet{\mathsf}{OML}{cmbr}{m}{it}

\newtheorem{theorem}{\bf Theorem}
\newtheorem{lemma}{\bf Lemma}

\newcommand{\bd}{\begin{description}}
\newcommand{\ed}{\end{description}}
\newcommand{\be}{\begin{enumerate}}
\newcommand{\ee}{\end{enumerate}}
\newcommand{\bi}{\begin{itemize}}
\newcommand{\ei}{\end{itemize}}
\newcommand{\bl}{\begin{list}}
\newcommand{\el}{\end{list}}
\newcommand{\bt}{\begin{tabbing}}
\newcommand{\et}{\end{tabbing}}

\begin{document}

{
\title{Age of Information in Locally Adaptive Frame Slotted ALOHA 
{\footnotesize}
\thanks{
The work of Z. Yue, H. H. Yang, and M. Zhang was supported in part by the National Natural Science Foundation of China under Grant 62271513, in part by the Zhejiang Provincial Natural Science Foundation of China under Grant LGJ22F010001, and in part by the Zhejiang – Singapore Innovation and AI Joint Research Lab. 
%(\textit{Corresponding Author: Howard H. Yang.}) 
The work of N. Pappas has been supported by the Swedish Research Council (VR), ELLIIT, and the European Union (ETHER, 101096526).}
}

\author{
      Zhiling~Yue$^\dagger$,
      Howard~H.~Yang$^\dagger$, 
      Meng~Zhang$^\dagger$,
      and Nikolaos Pappas$^\star$ \\
       
    $^\dagger$ \textit{ZJU-UIUC Institute, Zhejiang University, Haining 314400, China }\\
    $^\star$ \textit{Department of Computer and Information Science, Linköping University, Linköping 58183, Sweden}\\
    chi-ling@zju.edu.cn,\{haoyang, mengzhang\}@intl.zju.edu.cn, nikolaos.pappas@liu.se
}
    %~~~~~~~~~~~~~~~~~~~~~~~~~~~%
    %      Title footnote       %
    %~~~~~~~~~~~~~~~~~~~~~~~~~~~%
%         \thanks{ ($^*$: Equal contribution; \textit{Corresponding Author: Howard H. Yang.)}
%         This work was supported in part by the National Natural Science Foundation of China under Grant 62271513. The work of N. Pappas has been supported in part by the Swedish Research Council (VR), ELLIIT, Zenith, and the European Union (ETHER, 101096526).
%         % \thanks{
%         % This work was supported in part by the Zhejiang Provincial Natural Science Foundation of China under Grant No. LGJ22F010001. ($^*$: Equal contribution; \textit{Corresponding Author: Howard H. Yang.
%         % }
% %    \thanks{Manuscript submitted Mar, 2007, and revised Sep, 2007.}
% %    \thanks{
% %    This research was supported, in part, by
% %   }
% %    \thanks{
% %        H.\ Yang, J.\ Lee and T.\ Q.\ Quek are with
% %       the Singapore University of Technology and Design,
% %       20 Dover Drive, Singapore 138682
% %       (e-mail:{\,\,\,}\texttt{hao{\_}yang@mymail.sutd.edu.sg},{\,\,\,\,}\texttt{jmnlee@ieee.org}, \texttt{tonyquek@sutd.edu.sg}).
% %}

% }

  % \vspace{-0.5cm}

\maketitle
\acresetall
\thispagestyle{empty}
% \vspace{-0.5cm}
\begin{abstract}
We consider a random access network consisting of source-destination pairs. Each source node generates status updates and transmits this information to its intended destination over a shared spectrum. The goal is to minimize the network-wide Age of Information (AoI). We develop a frame slotted ALOHA (FSA)-based policy for generating and transmitting status updates, where the frame size of each source node is adjusted according to its local environment. The proposed policy is of low complexity and can be implemented in a distributed manner.
Additionally, it significantly improves the network AoI performance by ($a$) equalizing the update generation intervals at each source and ($b$) reducing interference across the network. Furthermore, we derive an analytical expression for the average network AoI attained for that policy. We evaluate the performance of the proposed scheme through simulations, which demonstrate that the locally adaptive FSA policy achieves a remarkable gain in terms of AoI compared to the slotted ALOHA counterpart, confirming the effectiveness of the proposed method.
% We consider a frame-slotted ALOHA (FSA)-based status updating protocol for a random access network, where the source-destination pairs constitute a Poisson bipolar deployment.To mitigate the harmful consequence of severe interference on geographically proximate links, we develop a decentralized status updating policy, i.e., sources can adjust their framesizes and updating rates by utilizing their local observations. With the goal of minimizing the network-wide age of information (AoI), we derive mathematical expressions for the suboptimal update policy at each source as well as the network average AoI.
% Simulation explores the performance of the proposed scheme and verifies the improvement of local information on AoI under slotted ALOHA (SA) protocol and the further optimization of the performance after expanding it to FSA.
\end{abstract}

\acresetall
%%%%%%%%%%%%%%%%%%%%%%%%%%%%%%%%%%%%%%%%%%%%%%%%%%%%
\section{Introduction}\label{sec:intro}
%\subsection{Motivation}
The rapid development of wireless communication and Internet-of-Things (IoT) technology has given rise to numerous real-time applications, including autonomous driving \cite{autoAoI}, healthcare system \cite{2022health}, and real-time human-AI interaction. 
% These latency-sensitive applications have a stringent demand for the timeliness of information since 
% outdated data may lead to wrong decisions, incurring potentially dangerous consequences \cite{2017Radio}.
These applications are highly sensitive to latency and require timely information, as outdated data can result in erroneous decisions, leading to potentially dangerous consequences \cite{2017Radio}.
In response to this concern, researchers have proposed the \textit{Age of Information }(AoI) metric to assess the timeliness of the received information \cite{KosPapAng:17, sunmodiano2019age, pap_2023}.
Unlike traditional metrics such as delay or throughput, AoI is receiver-oriented and measures the time elapsed since the latest received update was generated. 
Thus, unlike the source-centric metrics, AoI captures the ``freshness" of information. 
A broad array of research has been conducted on analyzing and designing techniques to optimize AoI under different systems.

The study of AoI originated in \cite{2012Real}, where the authors analyzed the average AoI under a standard fist-come-first-serve queue. 
The findings revealed that schemes for optimizing the conventional metrics may not effectively minimize AoI.
This discovery led to further exploration in subsequent works  \cite{2020Opti,PappasICC15,M2016On}, wherein various scheduling policies and packet management schemes were developed with a primary focus on point-to-point systems.
% Building upon the goal of minimizing AoI, subsequent works \cite{2020Opti,PappasICC15,M2016On} developed different scheduling policies and packet management schemes, primarily focusing on point-to-point systems. 

% a series of works extend \cite{2012Real} to design different scheduling policies \cite{2020Opti,HeYuaEph:16} or packet management schemes \cite{,PappasICC15,CosCodEph:16}.
% However, these works primarily focused on point-to-point systems.

In wireless networks, communications often occur over a shared spectrum, resulting in potential severe interference that significantly degrades communication quality or even leads to transmission failure.
In such a context, the signal-to-interference-plus-noise ratio (SINR) model \cite{2017Birth} is commonly used to characterize interference, which encompasses key properties of a wireless network, including fading, path loss, and co-channel interference.
Under this model, if the SINR received by the destination surpasses a predetermined threshold, the transmission is deemed successful; otherwise, it is considered failed.
The SINR model has promoted several studies on AoI in random access networks, which employed stochastic geometry to model the complex spatial deployments as point processes, significantly facilitating the interference analysis. 
For example, \cite{HuZhoZha:18} derived the upper and lower bounds on the average AoI distribution in Poisson networks by taking the locations of the interfering nodes into account.
Subsequently, \cite{2021SpatialDis} further improved this result 
% This result was improved in \cite{2021SpatialDis}, 
by deriving a tighter upper bound on the spatial distribution of the average peak AoI. 
In addition to these bounds, \cite{2021Understand} established a theoretical framework capturing the intricate spatiotemporal entanglement amongst transmitters in the network and provided analytical expressions for peak and average AoI. 
In \cite{MankarTWC21}, a stochastic geometric analysis of throughput and AoI in a cellular IoT network was presented.
% established a theoretical framework that captures the complex spatiotemporal entanglement of the network. It also provided analytical expressions for the peak and average AoI.
% Since the interference to each source is dominated by its spatially adjacent sources, 
Recognizing that the interference experienced by each source is primarily influenced by its spatially adjacent neighbors,
\cite{YanArafaQue:19} proposed a decentralized channel access strategy, enabling each source to set its channel access probability based on the observed surrounding information.
This scheme significantly reduced the peak AoI.
Following similar considerations, \cite{2021powctl} and \cite{2022Locally} leveraged the locally observed information to adjust each source node's transmit power and updating rate, respectively, to optimize the network AoI.

The aforementioned works utilized the slotted ALOHA (SA) protocol. Building upon this, \cite{2023FSA} proposed a frame SA (FSA) protocol that significantly enhances the AoI performance.
However, \cite{2023FSA} adopted a universally fixed framesize throughout the entire network, ignoring the discrepancy amongst interference levels at different spatial locations. In this paper, we improve the FSA-based protocol by developing a strategy that adapts the framesize and update rate of each node based on its local observation, improving the network AoI performance. 
Specifically, we model the source-destination pair as a Poisson bipolar network. 
Time is divided into slots of equal duration, and each source combines several consecutive time slots into a single frame. 
At the beginning of each frame, sources will decide with different probabilities whether to be active in this frame, where both the framesize and the probability of being active depend on their local observations. 
If a source decides to be active, it will pick one time slot in this frame uniformly at random, generate an updated sample, and transmit the information to the destination immediately; otherwise, it will stay silent for the entire duration of the frame. 
We employ the SINR model to quantify the quality of information delivery, where transmission is successful only when the SINR exceeds a given decoding threshold.
Our main contributions are summarized below.
\begin{itemize}
    \item We develop a locally adaptive FSA protocol that exploits the observed information at each node to set its framesize to minimize the AoI across the network. The proposed protocol is decentralized and has low implementation complexity. 
    \item We develop an analytical framework to evaluate the network average AoI. Our framework takes into account several attributes of a wireless system, including the deployment density, random activation of transmitters, channel gain, path loss, and interference.
    \item  The network average AoI can be significantly reduced by incorporating local observations into the policy design, whereas increasing the information about the surrounding environment will bring about further improvement.
\end{itemize}

%%%%%%%%%%%%%%%%%%%%%%%%%%%%%%%%%%%%%%%%%%%%%%%%%%%%
\section{System Model}\label{sec:sysmod}
In this section, we introduce the system model and the performance metric.
We also elaborate on the concepts of stopping sets and locally adaptive policies.
\subsection{Network Model}
\subsubsection{Spatial Deployment}
\begin{figure}[t!] 
  \centering{}

    {\includegraphics[width=0.94\columnwidth]{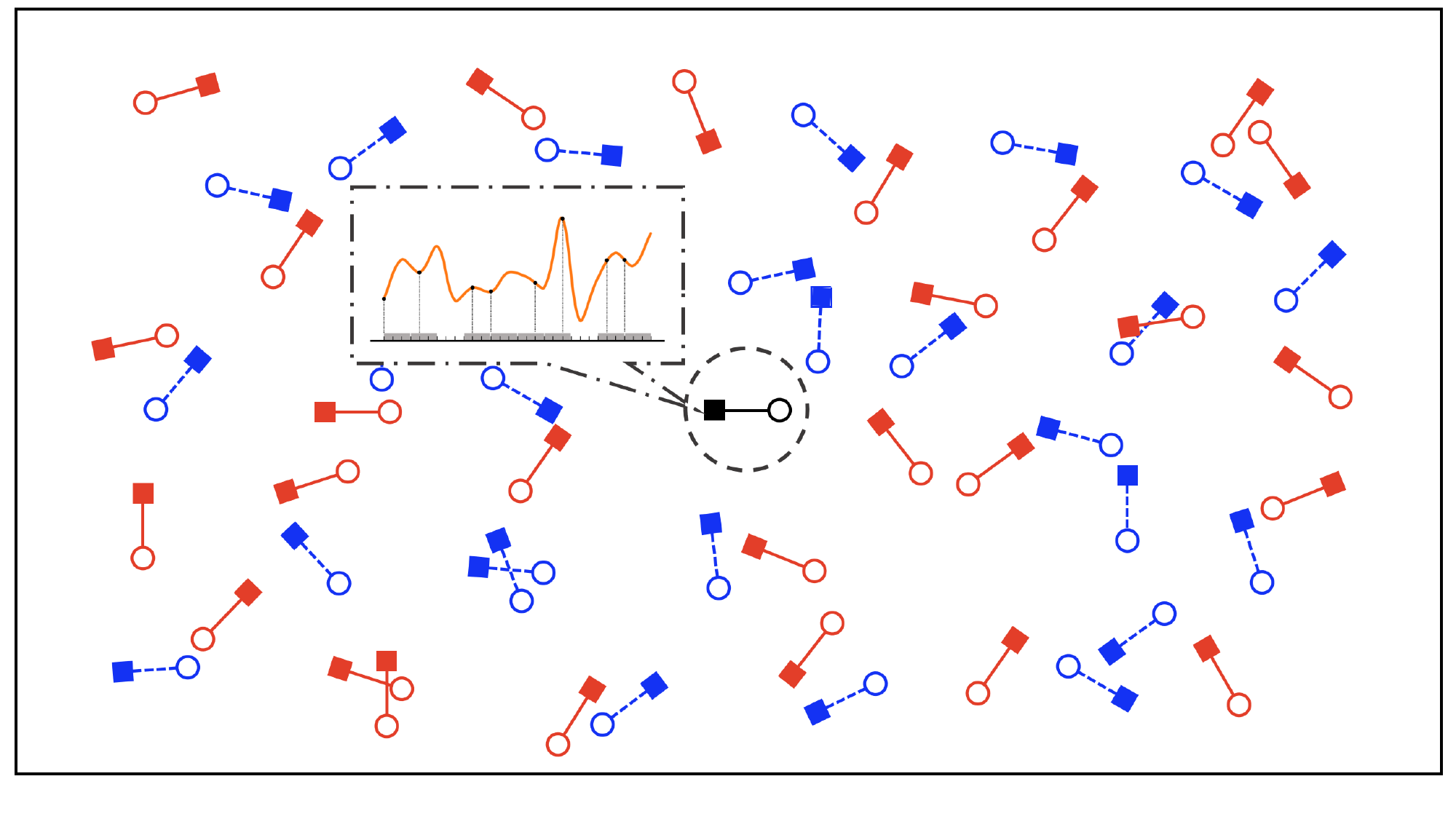}}

  \caption{A snapshot of the employed network model. The solid squares represent the source nodes that sample a physical process's most recent status information and transmit them to the destinations denoted by the circles. The solid black line is the typical link, the solid red lines represent other active links, and the dashed blue lines are inactive.}
  \label{fig:network}
\end{figure}
We consider a Poisson bipolar network on the Euclidean plane, as depicted in Figure~\ref{fig:network}.
%, constituted by source-destination pairs.
Specifically, the sources are deployed according to a homogeneous Poisson point process (PPP), denoted by $\Phi_{\mathrm{s}} = \{ X_i \}_{i=0}^\infty$, with spatial density $\lambda$.
Each source node has an exclusive destination located at a constant distance $r$ from it in a random orientation.
According to the displacement theorem \cite{BacBla:09}, the spatial distribution of destinations, denoted by $\Phi_{\mathrm{d}} = \{ y_i \}_{ i = 0 }^\infty$, is also subject to a homogeneous PPP with the same spatial density.
In this network, every source monitors a physical process and updates the status of its observation to the destination by sending a sequence of information packets. 
% Then, the resulting updated information is encapsulated into packets and sent to the destination
We assume that each node transmits at the same power $P_{ \mathrm{tx} }$\footnote{This assumption is made to facilitate the analysis. Note that the framework in this paper can be extended to design power control strategies as per \cite{2021powctl}.}, 
whereas the transmissions take place over a shared spectrum.
The signal propagation is affected by small-scale Rayleigh fading with a unit mean and path loss that obeys the power law. 
The channel fading varies independently across time and space.
The transmissions are also subject to white Gaussian thermal noise with variance $\sigma^2$.

\subsubsection{Temporal Pattern}
We assume that the network is synchronized. We segment the time into slots of the same duration, equal to the period required to complete a packet transmission. 
We adopt a \textit{generate-at-will} approach based on the FSA protocol for status updating \cite{2023FSA}. 
Specifically, for link $i$, we group a number of consecutive time slots into a frame of size $F_i$.
At the beginning of each frame, source $i$ independently decides whether to sample information in that frame with probability $\eta_i$ (also known as the update rate). 
If node $i$ decides to update, it selects a time slot in frame $F_i$ uniformly at random to generate the update information and sends it to the destination immediately. 
% Information generation occurs at the beginning of the time slot and the transmission occupies one time slot. 
At the end of the same time slot, if the received SINR exceeds a decoding threshold $\theta$, the information is successfully delivered. 
% In order to always transmit the most "fresh" information, each packet is always transmitted only once, that is, no feedback and retransmission mechanisms are used whether the transmission is successful or not.

Since the time scale of signal transmission and fading is much smaller than that of spatial dynamics, we assume the network is static; namely, the nodes are initially scattered randomly and remain constant in subsequent time slots.

\subsection{Performance Metric}
\begin{figure}[t!] 
  \centering{}

    {\includegraphics[width=0.94\columnwidth]{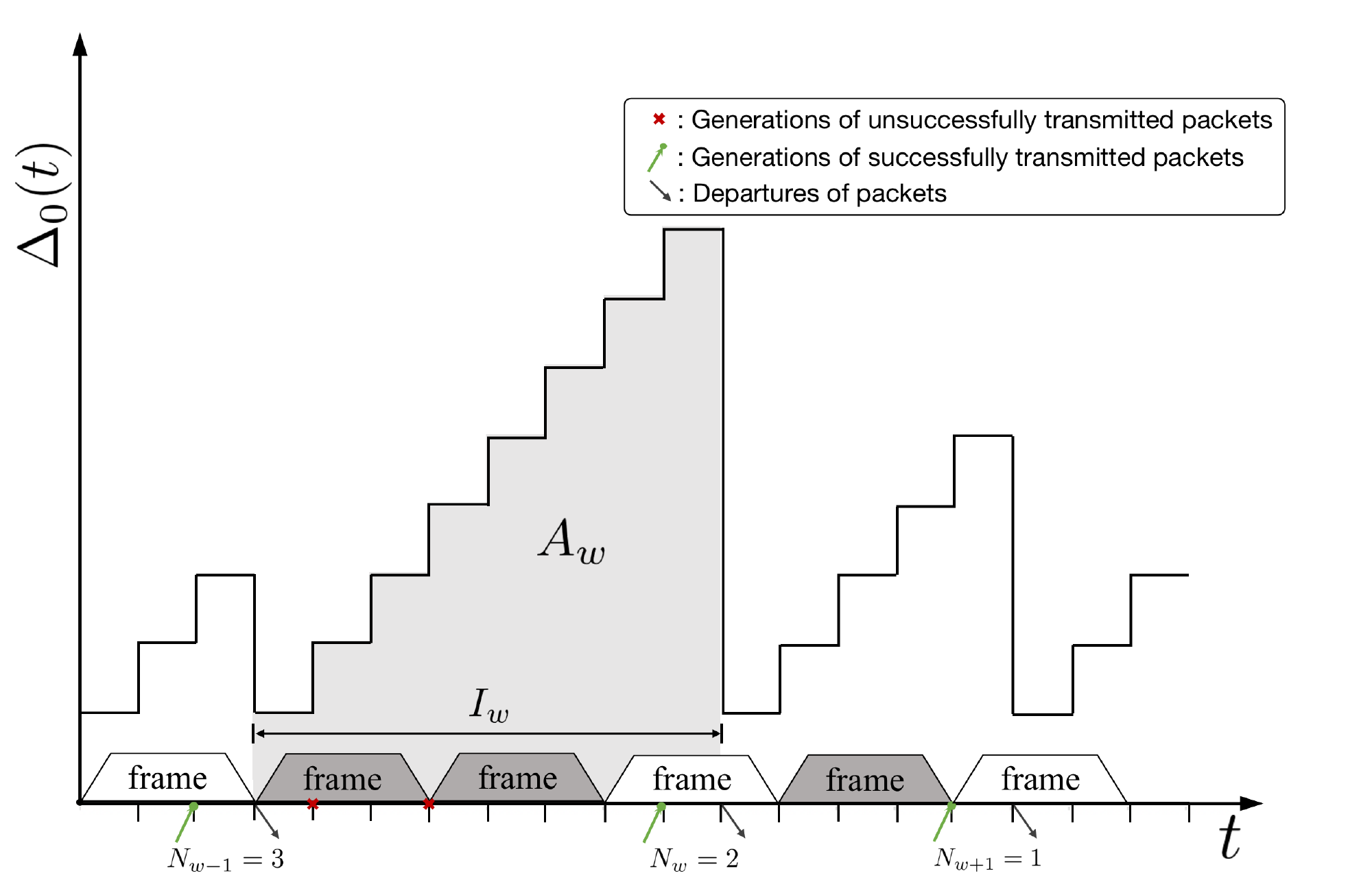}}

  \caption{An example of the AoI evolution over the typical link under the FSA status updating protocol. The framesize of the typical link is set as $F_0 = 3$.}
  \label{fig:FSA_AoI}
\end{figure}
This paper focuses on AoI, which quantifies the timeliness of information delivered in the network. 
As illustrated in Figure~\ref{fig:FSA_AoI}, AoI increases linearly\footnote{The framework of this paper can also be extended to the case of non-linear growth according to \cite{2022nonlinear}.} over time until the information at the destination is successfully updated. 
Without loss of generality, we choose an arbitrary link in the network as our \textit{typical link} and mark its receiver position as the origin. 
According to Slivnyak’s theorem\cite{BacBla:09}, the performance of this link is statistically identical to that of other links.
Thus, it can be used as a representative.
A formal expression for the instantaneous AoI over the typical link can then be written by
\begin{align}
    \Delta_0 (t) = t - G_0(t),
\end{align}
where $G_0(t)$ represents the generation time-stamp of the latest packet received by the typical destination at time $t$.

To capture the overall timeliness in the delivery of status updates through the typical link, we then define the time-average AoI at the typical receiver as following
\begin{align}
    \bar{\Delta}_0 = \lim_{T \to \infty }{\frac{1}{T}\sum_{t=1}^T \Delta_0(t)}.
\end{align}
Note that different nodes have distinct update strategies and are subject to different interference levels.
Therefore, $\bar{\Delta}_i$ is still a random variable.
As such, we average the time-average AoI of all the links over space and term the resultant quantity the \textit{network average AoI}\cite{2021Understand}.
Formally, it is given by
\begin{align}
    \bar{\Delta} &= \limsup_{R \to \infty} \frac{\sum_{i: X_i \in B(0,R)} \bar{\Delta}_i}{\lambda \pi R^2} = \mathbb{E} \left[ \bar{\Delta}_0 \right],
\end{align}
where $B(0,R)$ represents a disk with radius $R$ centered at the origin.
% where (a) follows from Campbell's theorem \cite{BacBla:09}, and $\mathbb{E}^0[\cdot]$ represents taking expectation with respect to the Palm distribution $\mathbb{P}^0$ of the stationary point process in which under $\mathbb{P}^0$ almost surely there is a node located at the origin \cite{BacBla:09}.

\subsection{Stopping Sets and Locally Adaptive Protocol}
Due to the spectrum's shared nature, simultaneous transmissions of wireless links in proximity can result in severe interference that hinders information delivery. 
On the other hand, if every source could obtain certain information about its nearby interferers, it can utilize that information to adjust the status update policy, balancing the timeliness of samples at the source and interference level across the network. 
As a result, the nodes may jointly minimize the network average AoI in a distributed but cooperative manner.

Thus, we configure sensors for each source to perceive the geographic information of the surrounding transmitters.
Then according to this knowledge, sources will adjust their framesizes and update rates to optimize the AoI performance of the network. Although the sensing capability of each source node is often limited, considering that concurrent transmissions in geographic proximity usually dominate interference, the local observations and the subsequent design are still meaningful. 
We model the observation window by the notion of \textit{stopping set}\cite{YanArafaQue:19, 2014Analy}.
Specifically, given the spatial deployment $\Phi \triangleq {\Phi_{ \mathrm{s} } } \cup {\Phi_{ \mathrm{d} } }$, the stopping set $W = W(\Phi)$ is a measurable function that maps $\Phi$ to a Borel set $\mathcal{H}$ in $\mathbb{R}^2$.
One can determine whether the event $\{W \subset \mathcal{H} \}$ occurs or not just with the knowledge about points of $\Phi$ in $\mathcal{H}$.
For clarity, we choose a generic source $i$, 
% and denote its location as $X_i \in \Phi_{ \mathrm{s} }$ and its dedicated destination location denoted as $y_i \in \Phi_{ \mathrm{d} }$.
and demonstrate two types of stopping set construction as examples:
\subsubsection{Random Stopping Sets}
Given a compact set $\mathcal{H}$, try constructing a minimal disk centered on the source $i$ and containing the nearest $p$ nodes around it. 
We define the radius of this disk as $R_p = \min \{R \geq 0: \Phi(B(X_0,R))=p \}$. 
If the disk exists, we know that $B(X_0,R_p) \subset \mathcal{H}$, otherwise $B(X_0,R_p) \not\subset \mathcal{H}$.
\subsubsection{Deterministic Stopping Sets}
Similar to the above, given a predetermined radius $R$ and a compact set $\mathcal{H}$, we construct a disk centered on $X_i$ and increase its area.
We will stop increasing until the moment when either (a) the radius of this disk reaches the value $R$ or (b) the disk grows large enough to touch the complement of $\mathcal{H}$.
%Similar to the above, given $\mathcal{H}$, centered on $X_i$ and without hitting the complement of $\mathcal{H}$, we
%increase the area of the disk until its radius reaches a predetermined value $R$.
If the former happens, we know that $B(X_0,R) \subset \mathcal{H}$; but if the latter happens, we know that $B(X_0,R) \not\subset \mathcal{H}$.
%If such a disk exists, we know that $B(X_0,R) \subset \mathcal{H}$; otherwise, we stop increasing and know $B(X_0,R) \not\subset \mathcal{H}$ as soon as the disk grows large enough to touch the complement of $\mathcal{H}$.
Either way, we do not need information outside the set $\mathcal{H}$; hence the stopping set is respect to $\Phi$. 

Based on this concept, we can define the locally adaptive update rate and framesize at source node $i$ as following
\begin{align}
    \eta_i = \eta(W(S_{X_i}(\Phi))),\\
    F_i = F(W(S_{X_i}(\Phi))),
\end{align}
where $S_x$ is a shift operation that converts the locations of nodes in any set $\mathcal{H} \subset \mathbb{R}^2$ into a coordinate representation with $x$ as the origin (i.e. $S_x(\mathcal{H}) = \{z-x: z \in \mathcal{H}\}$), $\eta: \mathbb{R}^2 \to [0,1]$ is a measurable function that maps the local information $W(S_{X_i}(\Phi))$ into an update rate, and $F: \mathbb{R}^2 \to \mathbb{N}^*$ is a measurable function that intakes $W(S_{X_i}(\Phi))$ and produces a frame size.
Two points are noteworthy:
i) both $\eta_i$ and $F_i$ are functions of nearby geometry, not their actions.
Therefore, given the observation window, they are independent of each other; 
ii) once the policy (i.e., the pair $(\eta, F)$ of update rate and framesize) is determined, it will not change over time.

%========================================%
%            Analysis
%========================================%
\section{Design and Analysis} \label{Sec:AveAoI_Analysis}
In this section, we develop a status update policy that enables the source nodes to adjust their framesizes in accordance with their local observation, which improves the AoI performance of the network. 
We also analyze the network average AoI under such a policy.

\subsection{Preliminaries}
We start with analyzing the SINR statistics, which quantify the quality of the wireless links. 
Specifically, if the typical source sends out an information packet at time $t$, the received SINR can be written by
\begin{align} \label{equ:SINR}
    \textrm{SINR}_{0,t} = \frac{P_{ \mathrm{tx} } h_{0,t} r^{-\alpha}}{\sum_{i \neq 0}P_{ \mathrm{tx} } h_{i,t} v_{i,t} \Vert  X_i  \Vert ^{-\alpha}+\sigma^2},
\end{align}
where $h_{i,t} \sim \exp(1)$ denotes the channel fading between source $i$ and the typical receiver, $\alpha$ stands for the path loss exponent, $v_{i,t}$ indicates whether source $i$ is activated at $t$ (in this case, $v_{i,t}=1$) or not (in this case, $v_{i,t}=0$), and $\Vert \cdot \Vert$ denotes the Euclidean norm.

Because the network is static, we characterize the rate of information delivery, which is directly affected by the SINR,
using the \textit{conditional transmission success probability}\cite{2015The}.
To be more precise, the conditional transmission success probability over the typical link represents the probability of the received SINR at the typical receiver exceeding a decoding threshold $\theta$, given the point process $\Phi$,
namely\footnote{Since the considered point process in this work is stationary, we drop the subscript $t$ in the sequel.},
\begin{align} \label{equ:CndTXProb}
\mu_0^\Phi = \mathbb{P}(\mathrm{SINR_0}>\theta \mid \Phi).
\end{align}
We first average out the effects of channel fading in a similar way to \cite{2023FSA} and arrive at the analytical expression of $\mu_0^\Phi$.
\begin{lemma} \label{lem:mu_0}
\textit{Given the point process $\Phi$, the conditional transmission success probability over the typical link is
\begin{align} \label{equ:CndTXSuccProb}
    \mu_0^{\Phi} = e^{-\frac{\theta r^\alpha}{\rho}} \prod_{i \neq 0}\left(  1-\frac{\eta_i /F_i}{1+\Vert X_i \Vert ^{\alpha}/\theta r^\alpha} \right),
\end{align}
where $\rho = P_{ \mathrm{tx} }/\sigma^2$ denotes the signal-to-noise ratio (SNR).
}
\end{lemma}
\begin{IEEEproof}
%The proof is similar to Lemma~1 in \cite{2023FSA} and hence omitted here.
According to our transmission protocol, we note that for any source $i$, it decides to update in a typical frame independently with probability $\eta_i$, and if it decides, it randomly selects one time slot in this frame according to a uniform distribution.
Consequently, at any given time slot, source $i$ activates with probability $\eta_i/F_i$, i.e., $\mathbb{P}(\nu_i=1)=\frac{\eta_i}{F_i}$.
Then, we can derive Lemma~\ref{lem:mu_0} as following
\begin{align}
    \mu_0^\Phi &= \mathbb{P} \left(\frac{P_{ \mathrm{tx} }h_0r^{-\alpha}}{\sum_{i \neq 0}P_{ \mathrm{tx} }h_i\nu_i\parallel \!\! x_i \!\! \parallel ^{-\alpha}+\sigma^2} >\theta \Big\vert \Phi \right)
    \nonumber\\
    &= \mathbb{P} \left(h_0 > \theta r^\alpha\sum_{i \neq 0}h_i \nu_i\parallel \!\! x_i \!\! \parallel ^{-\alpha} + \frac{\theta \sigma^2}{P_{ \mathrm{tx} }}  \Big\vert \Phi \right)
    %\nonumber\\
    %&= \mathbb{E} \left[ \exp \Big(-\theta r^\alpha \sum_{i \neq 0}h_i \nu_i\parallel \!\! x_i \!\! \parallel ^{-\alpha} \Big)\Big\vert \Phi \right]
     \nonumber\\
     &=e^{-\frac{\theta r^\alpha}{\rho}} \mathbb{E} \left[ \prod_{i \neq 0}\exp\left(-\theta r^\alpha h_i \nu_i\parallel \!\! x_i \!\! \parallel ^{-\alpha}\right) \Big\vert \Phi \right]
    \nonumber\\
     &\stackrel{(a)}{=}e^{-\frac{\theta r^\alpha}{\rho}} \mathbb{E} \left[ \prod_{i \neq 0} \frac{1}{1+\theta r^\alpha \nu_i \parallel \!\! x_i \!\! \parallel ^{-\alpha}} \right]
    \nonumber\\
     &\stackrel{(b)}{=}e^{-\frac{\theta r^\alpha}{\rho}}  \prod_{i \neq 0}\left(\frac{\eta_i/F_i}{1+\theta r^\alpha \parallel \!\! x_i \!\! \parallel ^{-\alpha}}+1-\frac{\eta_i}{F_i} \right),
     % \nonumber\\
     % &= e^{-\frac{\theta r^\alpha}{\rho}} \prod_{i \neq 0}\left(  1-\frac{\eta_i /F_i}{1+\Vert X_i \Vert ^{\alpha}/\theta r^\alpha} \right),
\end{align}
where (a) follows since $\{ h_i \}_{i=1}^\infty$ are i.i.d. random variables following the exponential distribution with unit mean, and (b) follows as $\{ \nu_i \}_{i=1}^\infty$ are independent of each other.
\end{IEEEproof}

Next, we adopt a graphical method to calculate the conditional time-average AoI over the typical link. 
\begin{lemma} \label{lem:con_AoI}
\textit{Conditioned on the spatial topology $\Phi$, the time-average AoI over the typical link is 
\begin{align} \label{equ:con_AoI}
    \bar{\Delta}_0 =\frac{F_0^2-1}{12F_0}\times \eta_0 \mu_0^\Phi + \frac{F_0}{\eta_0 \mu_0^\Phi}+\frac{1-F_0}{2}.
    %\bar{\Delta}_0=\frac{F^2-1}{12F}\times \eta \mathbb{E}(\mu_0^\Phi)+\frac{F}{\eta} \times \mathbb{E}(\frac{1}{\mu_0^\Phi})+\frac{1-F}{2}.
\end{align}
}
\end{lemma}
\begin{IEEEproof}
The proof is similar to Theorem~1 in \cite{2023FSA} thus omitted.
\end{IEEEproof}

Note that if all the nodes are updating in each frame and every information packet can be successfully delivered upon each transmission attempt, the network average AoI achieves its minimum.
This ideal scenario offers a fundamental lower bound for the network AoI performance as
\begin{align} \label{equ:low_bound}
    \bar{\Delta}_0 \geq \frac{F_0^2-1}{12F_0} + F_0 + \frac{ 1 - F_0 }{2} = \frac{ 7 F_0 }{12} - \frac{1}{12F_0} +\frac{1}{2}.
\end{align}
In consequence, we have
\begin{align}
\bar{\Delta}_0 \sim \frac{7}{12}F_0, \quad F_0 \rightarrow \infty
\end{align}
indicating that for large $F$, the network AoI increases monotonically with the frame size.
Therefore, it is important to adequately adjust the framesize to achieve an optimal age performance. 

\subsection{Status Update Policy}
From Lemma~\ref{lem:mu_0} and Lemma~\ref{lem:con_AoI}, we notice that the update rate and framesize jointly influence the network average AoI, where the effects are explicitly reflected by \eqref{equ:con_AoI} and implicitly contained in \eqref{equ:CndTXSuccProb}.
% we notice that the network average AoI is related to the updating rate and the framesize of the adopted FSA protocol.
%By deconditioning on the $\mu_0^\Phi$, we can obtain a more analytical expression for the network average AoI when 
As such, we can cast the policy design into the following optimization problem: given the observation window $W$, find the update rate-frame size pair $(\eta, F)$ such that
\vspace{-0.2cm}
\begin{align} \label{pblm:Vanilla_FSA_Opt}
    \min_{ \eta, F } \quad &\mathbb{E} \left[\bar{\Delta}_0(\eta_0, F_0) \vert W \right]
    \\
    \mathrm{s. t.} \quad&0 \leq \eta_i = \eta(W(S_{X_i}(\Phi))) \leq1, 
    \nonumber\\
    &F_i = F(W(S_{X_i}(\Phi))) \in \mathbb{N}^*, \qquad \forall i \in \mathbb{N}.
    % \nonumber
\end{align}

In order to solve \eqref{pblm:Vanilla_FSA_Opt}, we first derive an analytical expression for $\mathbb{E}[\bar{\Delta}_0(\eta_0, F_0) \vert W ]$.
Specifically, when the observation window is given by a stopping set $W = W( \Phi )$, we can decondition $\mu_0^\Phi$ in \eqref{equ:con_AoI} over $\Phi$, which results in the following expression for the network AoI.
\begin{lemma} \label{lem:AoI}
\textit{Conditioned on the stopping set $W$, the network average AoI is given by
\begin{align} \label{equ:AoI_W}
    &\mathbb{E} \left[\bar{\Delta}_0 \vert W \right]=\mathbb{E} \left[\frac{F_0}{\eta_0}\exp\!{\Big(\frac{\theta r^\alpha}{\rho}+\frac{C\eta_0}{F_0} \big(1\!-\!\frac{\eta_0}{F_0}\big)^{\delta-1}}\Big) \Big\vert W \right]
    \nonumber\\
    &+ \mathbb{E}\left[\frac{(F_0^2\!-\!1)\eta_0}{12F_0}\exp\!{\Big(\!-\! \frac{\theta r^\alpha}{\rho}\!-\!\frac{C\eta_0}{F_0}\Big)} \Big\vert W \right] + \frac{1\!-\! \mathbb{E}[F_0 \vert W]}{2},
    %\bar{\Delta}_0=\frac{F^2-1}{12F}\times \eta \mathbb{E}(\mu_0^\Phi)+\frac{F}{\eta} \times \mathbb{E}(\frac{1}{\mu_0^\Phi})+\frac{1-F}{2}.
\end{align}
where $\delta=2/\alpha$, $C=\lambda \pi r^2 \theta^\delta \Gamma(1-\delta)\Gamma(1+\delta)$, and $\Gamma(\cdot)$ is the Gamma function \cite{2015The}.
}
\end{lemma}
\begin{IEEEproof}
Please see Appendix~\ref{pro:lem3}.
\end{IEEEproof}

From this result we have the following observations. 

% Due to the complexity of this joint optimization problem, we propose the following two observations.

\textbf{Observation 1:}\textit{ Given any updating policy ($\eta, F$),  we can always transform it into another one as ($1, \tilde{F}$), where $\tilde{F} = F/\eta$, and obtain a smaller network average AoI.
}

This statement can be formally justified by comparing the network average AoI attained under the two policies as follows
\begin{align}
    &\mathbb{E} \left[\bar{\Delta}_0\Big( 1, \frac{F_0}{\eta_0} \Big) \Big\vert W \right] - \mathbb{E} \left[\bar{\Delta}_0(\eta_0, F_0) \vert W \right]
    \nonumber\\
    &= \mathbb{E}\left[\frac{(F_0/\eta_0)^2\!-\!{F_0}^2}{12F_0/\eta_0}\exp\!{\Big(\!-\frac{\theta r^\alpha}{\rho}\!-\!\frac{C\eta_0}{F_0}\Big)} +\frac{F_0 \!-\! \frac{F_0}{\eta_0}}{2}  \Big\vert W \right]
    \nonumber\\
    &< \mathbb{E}\left[\frac{\frac{F_0}{\eta_0}-\!F_0 \eta_0}{12} +\frac{F_0 \!-\! \frac{F_0}{\eta_0}}{2}  \Big\vert W \right]
    \nonumber\\
    &= \mathbb{E}\left[ \frac{F_0 ( 1 - \eta_0 )( \eta_0 - 5 )}{ 12 \eta_0 } \Big\vert W \right]<0.
\end{align}

Observation~1 also implies that having each node update in each frame is beneficial for the age performance of the network. 
This is because the frame structure improves network AoI by ($a$) reducing mutual interference amongst the transmitters and ($b$) equalizing the update generation intervals at each source, which is instrumental in reducing AoI. 
As a result, the optimization problem in \eqref{pblm:Vanilla_FSA_Opt} reduces to 
\begin{align} \label{pblm:Varnt_FSA_Opt}
    \min_{F} \quad\! &\mathbb{E} \bigg[ F_0 \exp{\bigg(\frac{\theta r^\alpha}{\rho} +  \frac{C}{F_0} \Big(  \frac{F_0 - 1}{F_0}  \Big)^{\delta  -  1}}  \bigg) 
    \nonumber\\
    & ~+ \frac{F_0^2 - 1}{12F_0}   \exp {\Big( - \frac{\theta r^\alpha}{\rho} - \frac{C}{F_0}  \Big)} + \frac{1 -  F_0}{2} \Big\vert W \!\bigg]
    \\
    \mathrm{s. t.} \quad\! &F_i = F(W(S_{X_i}(\Phi))) \in \mathbb{N}^*, \qquad \forall i \in \mathbb{N}.    
\end{align}
% By taking the derivation of $\bar{\Delta}$ with respect to $F$ and assigning it to zero, we can have the equation for the optimal framesize $F^*$ versus network parameters, which can be written as:
% \begin{align}
%     &\left[1 \!+\! \Big(1 \!-\! \frac{1}{F^*} \Big)^{\delta \!-\! 2} \frac{C(\delta \!-\! F^*)}{{F^*}^2} \right]\exp\!{\bigg(\! \frac{C}{F^*} \Big(\! \frac{F^*\!-\!1}{F^*} \!\Big)^{\delta \!-\! 1}} \!\bigg)
%     \nonumber\\
%     &+ \frac{{F^*}^3 \!+\! {F^*}^2 \!+\! F^* \!-\! 12}{12{F^*}^3} \exp{\Bigg(\!-\!\frac{C}{F^*} \Bigg)} -\frac{1}{2}=0
% \end{align}
% This equation can be solved efficiently with popular software such as Matlab.
% With this solution, we can obtain the optimal status update policy ($1,F^*$).

% From this observation, we have obtained the optimal framesize, but it is still not a closed-form solution.
% As this optimization problem remains hard to solve, we will give a suboptimal solution in the sequel.

\textbf{Observation 2: }\textit{The network average AoI under FSA is always smaller than that under SA.
In other words, if the solution to \eqref{pblm:Varnt_FSA_Opt} is to adopt $F_i$ at node $i$, then the network average AoI achieved by using SA with $\eta_i = \frac{1}{F_i}$ would be an upper bound to the optimal one.
}

This statement can be verified in a similar spirit as \cite{2023FSA}.

Armed with the above result, noticing that directly solving \eqref{pblm:Varnt_FSA_Opt} by finding the optimal framsize for each source node is difficult, we can opt for a suboptimal solution by minimizing the upper bound:
\begin{align} \label{pblm:SA_Opt}
    \min_{\eta} \quad &\mathbb{E} \left[\frac{1}{\eta_0 \mu_0^\Phi} \Big\vert W \right]
    \\ \label{equ:SA_Opt_cnsrt}
    \mathrm{s. t.} \quad&0 \leq \eta_i = \eta(W(S_{X_i}(\Phi))) \leq1, \quad \forall i \in \mathbb{N}    
\end{align}
where \eqref{pblm:SA_Opt} is the network average AoI obtained under the locally adaptive SA policy \cite{2022Locally}, and \eqref{equ:SA_Opt_cnsrt} confines the update rate to be within a feasible range. 

The following theorem provides a solution to the above optimization problem.
\begin{theorem} \label{the:opt_eta}
\textit{Given the observation window $W$, 
the solution to \eqref{pblm:SA_Opt} is obtained by solving a fixed-point equation at each node, where the one at source $i$ is specifically given as follows
\begin{align} \label{equ:opt_eta}
    \frac{1}{\eta} \!-\! \sum_{\substack{y_i \in W,\\ j \neq i}} \frac{1}{1 \!+\! D_{ij} \!-\! \eta} \!-\!\!\! \int\limits_{\mathbb{R}^2 \backslash W} \!\! \frac{\lambda(1 \!+\! \Vert x \Vert^\alpha/\theta r^\alpha) \mathrm{d}x}{(1 \!-\! \eta \!+\! \Vert x \Vert^\alpha/\theta r^\alpha)^2}=0,
\end{align}
if the following condition holds
\begin{align}
    \sum_{y_i \in W, j \neq i} \frac{1}{D_{ij}} + \lambda \int_{\mathbb{R}^2 \backslash W} \bigg(\frac{\theta r^\alpha}{\Vert x \Vert^\alpha} + \frac{\theta^2 r^{2\alpha}}{\Vert x \Vert^{2\alpha}} \bigg)\mathrm{d}x > 1,
\end{align}
where $D_{ij} = \Vert X_i-y_j \Vert^\alpha/\theta r^\alpha$; and set as $\eta=1$ otherwise.
}
\end{theorem}
\begin{IEEEproof}
    With the expression \eqref{equ:bipo_2}, we can decondition the spatial topology in \eqref{pblm:SA_Opt} and get the following
    \begin{align} \label{equ:AoI_SA_de}
        \mathbb{E} \left[\frac{1}{\eta_i \mu_i^\Phi} \Big\vert W \right] = \frac{ \exp \bigg( {\! \frac{\theta r^\alpha}{\rho} + \lambda \!\int_{x \in \mathbb{R}^2 \backslash W}\!\! \frac{\eta_i \quad \!\! \mathrm{d}x}{1-\eta_i + \!\Vert x \Vert ^\alpha/\theta r^\alpha}} \bigg) }{\eta_i \prod_{\substack{j \neq i,\\ y_j \in W}}\!\! \left(1\!-\!\frac{\eta_i}{1 \!+\! D_{ij}} \right)}.
    \end{align}
    
    In order to devise the optimal updating rate for minimizing the network average AoI under SA protocol, we obtain the derivative of \eqref{equ:AoI_SA_de} with respect to $\eta$, which yields the following
    \begin{align} \label{equ:der_AoI}
         \frac{\partial}{\partial \eta_i} \mathbb{E} \left[\frac{1}{\eta_i \mu_i^\Phi} \Big\vert W \right] = \varphi(\eta_i) \times \mathbb{E} \left[\frac{1}{\eta_i \mu_i^\Phi} \Big\vert W \right],
    \end{align}
    in which $\varphi(\eta_0)$ is derived as:
    \begin{align}
        \varphi(\eta_i) = \frac{1}{\eta_i} \!-\!\! \sum_{\substack{y_j \in W,\\ j \neq i}} \frac{1}{1 \!+\! D_{ij} \!-\! \eta} \!-\!\!\! \int\limits_{\mathbb{R}^2 \backslash W} \!\! \frac{\lambda(1 \!+\! \Vert x \Vert^\alpha/\theta r^\alpha) \mathrm{d}x}{(1 \!-\! \eta_i \!+\! \Vert x \Vert^\alpha/\theta r^\alpha)^2}.
    \end{align}
    
    Since $\mathbb{E} \left[\frac{1}{\eta \mu^\Phi} \Big\vert W \right]$ is the the network average AoI under SA protocol, it will always have a positive value.
    Note that, $\varphi(\eta)$ is a continuous and monotonically increasing function about $\eta$ over the interval $[0,1]$, and when $\eta \to 0$, we have $\varphi(\eta) \to - \infty$.
    Therefore, if $\varphi(1)>0$, $\frac{\partial}{\partial \eta} \mathbb{E} \left[\frac{1}{\eta \mu^\Phi} \Big\vert W \right] = 0$ will have and only have one unique solution that optimizes \eqref{pblm:SA_Opt}.
    However, if $\varphi(1)<0$, we can get that $\frac{\partial}{\partial \eta} \mathbb{E} \left[\frac{1}{\eta \mu^\Phi} \Big\vert W \right] < 0$ for all $\eta \in [0,1]$. 
    Thus the network average AoI under SA protocol monotonically decreases with $\eta$ and hence the optimized updating protocol takes at $\eta =1$.
\end{IEEEproof}

Note that Theorem~1 produces an update rate $\eta_i$ for node $i$ in the form of a function that intakes the spatial information contained in the stopping set $W$. 

Using the results given in this theorem, we construct a locally adaptive FSA protocol by setting the framesize of node $i$ as $\hat{F}_i = \lceil \frac{1}{\eta_i} \rceil$, where $\lceil \cdot \rceil$ denotes the ceiling function.
Albeit such an approach only offers a suboptimal solution to \eqref{pblm:Varnt_FSA_Opt}, \textit{it will be demonstrated in Section~IV that this suboptimal solution achieves a significant improvement in terms of the network average AoI}. 

\subsection{Network average AoI}
Employing the update policy designed above, we can calculate the network average AoI as follows
\begin{align}
    \bar{\Delta} = \sum_{l=1}^\infty \bar{\Delta}(1, l) \mathbb{P}(F^*=l), \quad l \in \mathbb{N}^*.
\end{align}
The problem's tricky part is finding the distribution of $F^*$. 
Based on Observation 2, we can derive the distribution of $\hat{F}$ under the suboptimal update policy as follows:
\begin{align}
    \mathbb{P}(\hat{F}=l) &= \mathbb{P} \left(
    \left\lceil \frac{1}{\eta} \right\rceil=l \right)
    \nonumber\\
    &= \mathbb{P} \left(\frac{1}{\eta}  \leq l < \frac{1}{\eta}+1 \right)
    \nonumber\\
    &= \mathbb{P} \left(\frac{1}{l}  \leq \eta < \frac{1}{l-1} \right)
    \nonumber\\
    &= \mathbb{P} \left(\eta \geq \frac{1}{l} \right)-\mathbb{P} \left(\eta \geq \frac{1}{l-1} \right).
\end{align}
When a deterministic stopping set $W$ is given, the complementary cumulative distribution function (CCDF) of updating rate $\eta$ can be derived according to the Theorem 2 in \cite{2022Locally} as
\begin{align}
\mathbb{P}(\eta > \kappa) =
     \frac{1}{2\pi}\!\!\int_{-\infty}^\infty \!\! \mathcal{L}_{\mathcal{U}(\kappa,W)}(j \omega)\frac{e^{j \omega(1\!-\! \mathcal{V}(\kappa,W))-1}}{j \omega} \mathrm{d}\omega,
\end{align}
where $0<\kappa<1$, and if $\kappa=1$, the following holds
\begin{align}
     \mathbb{P}\left(\mathcal{U}(1,W)<1-\mathcal{V}(1,W) \right).
\end{align}
Here, $j = \sqrt{-1}$, while the $\mathcal{U}(\kappa,W)$, $\mathcal{V}(\kappa,W)$, and the Laplace transform of $\mathcal{U}(\kappa,W)$ are respectively given as
\begin{align}
    \mathcal{U}(\kappa,W) &= \sum_{y_i \in \Phi_{\mathrm{d}}} \frac{\kappa \cdot \chi_{\{y_i \in W, i \neq 0 \}}}{\frac{\Vert y_i \Vert^\alpha}{\theta r^\alpha}+1-\kappa},\\
    \mathcal{V}(\kappa,W) &= \int_{\mathbb{R}^2 \backslash W} \frac{\lambda \kappa(1 \!+\! \Vert x \Vert^\alpha/\theta r^\alpha) \mathrm{d}x}{(1 \!-\! \kappa \!+\! \Vert x \Vert^\alpha/\theta r^\alpha)^2},\\
    \mathcal{L}_{\mathcal{U}(\kappa,W)}(s) &= \exp \!\left(\!-\! \lambda \!\! \int_W \!\! \Big[1 \!-\! \exp \!\Big(\!-\! \frac{s \kappa \theta r^\alpha}{\Vert x \Vert^\alpha \!\!+\!\! (1\!\!-\!\!\kappa)\theta r^\alpha} \Big)\!\Big] \mathrm{d}x \!\right).
\end{align}

%\subsubsection{No Local Observation}
If sources do not observe any available local information (i.e. $W = \emptyset$), following sights in Observation 1, we will configure a fixed size frame for each node and assign them to keep updating (i.e. $\eta=1$). In such a scheme, the network average AoI can be derived as
\begin{align} \label{equ:bipo_average}
    \bar{\Delta} \!=\! F\! \exp\!{\Big( \frac{\theta r^\alpha}{\rho}\!\!+\!\! \frac{C}{F} \big(1\!\!-\!\!\frac{1}{F}\big)^{\delta\!-\!1}}\!\Big)\!\!+\!\!\frac{F^2\!\!-\!\!1}{12F} \! \exp\!{\Big(\!\!-\!\!\frac{\theta r^\alpha}{\rho}\!\!-\!\!\frac{C}{F} \Big)}\!\!+\!\!\frac{1\!\!-\!\!F}{2},
\end{align}
By taking the derivative of $\bar{\Delta}$ with respect to $F$ and assigning it to zero, we can have the equation for the optimal framesize $F^*$ versus network parameters, which can be written as
\begin{align}
    &\left[1 \!+\! \Big(1 \!-\! \frac{1}{F^*} \Big)^{\delta \!-\! 2} \frac{C(\delta \!-\! F^*)}{{F^*}^2} \right]\exp\!{\bigg(\!\frac{\theta r^\alpha}{\rho}\!+\! \frac{C}{F^*} \Big(\! \frac{F^*\!-\!1}{F^*} \!\Big)^{\delta \!-\! 1}} \!\bigg)
    \nonumber\\
    &+ \frac{{F^*}^3 \!+\! {F^*}^2 \!+\! F^* \!-\! 12}{12{F^*}^3} \exp{\Bigg(\!-\!\frac{\theta r^\alpha}{\rho}\!-\!\frac{C}{F^*} \Bigg)} -\frac{1}{2}=0
\end{align}
This equation can be solved numerically with popular software such as Matlab.
With this solution, we can obtain the optimal status update policy ($1,F^*$).

%========================================%
%     Simulation & Numerical Results
%========================================%
\section{Simulation and Numerical Results} \label{Sec:SimNum_Results}
In this section, we examine the performance of the proposed policy through simulations.
We scatter (on average) $1,000$ source-destination pairs according to a Poisson bipolar model with spatial density $\lambda$, in a region of area $\frac{1000}{\lambda}$.
We repeat this process $10$ times. 
We collect the statistics over $10,000$ time slots for every communication link in each simulation run to calculate the network average AoI.
Unless special statements, we use the following parameters: $\alpha=3.8$, $r = 30$, $\theta=0~\mathrm{dB}$, $P_{\mathrm{tx}}=23 ~\mathrm{dBm}$, and $\sigma^2 = -96~\mathrm{dBm}$.

\begin{figure}[t!] 
  \centering{}

    {\includegraphics[width=0.9\columnwidth]{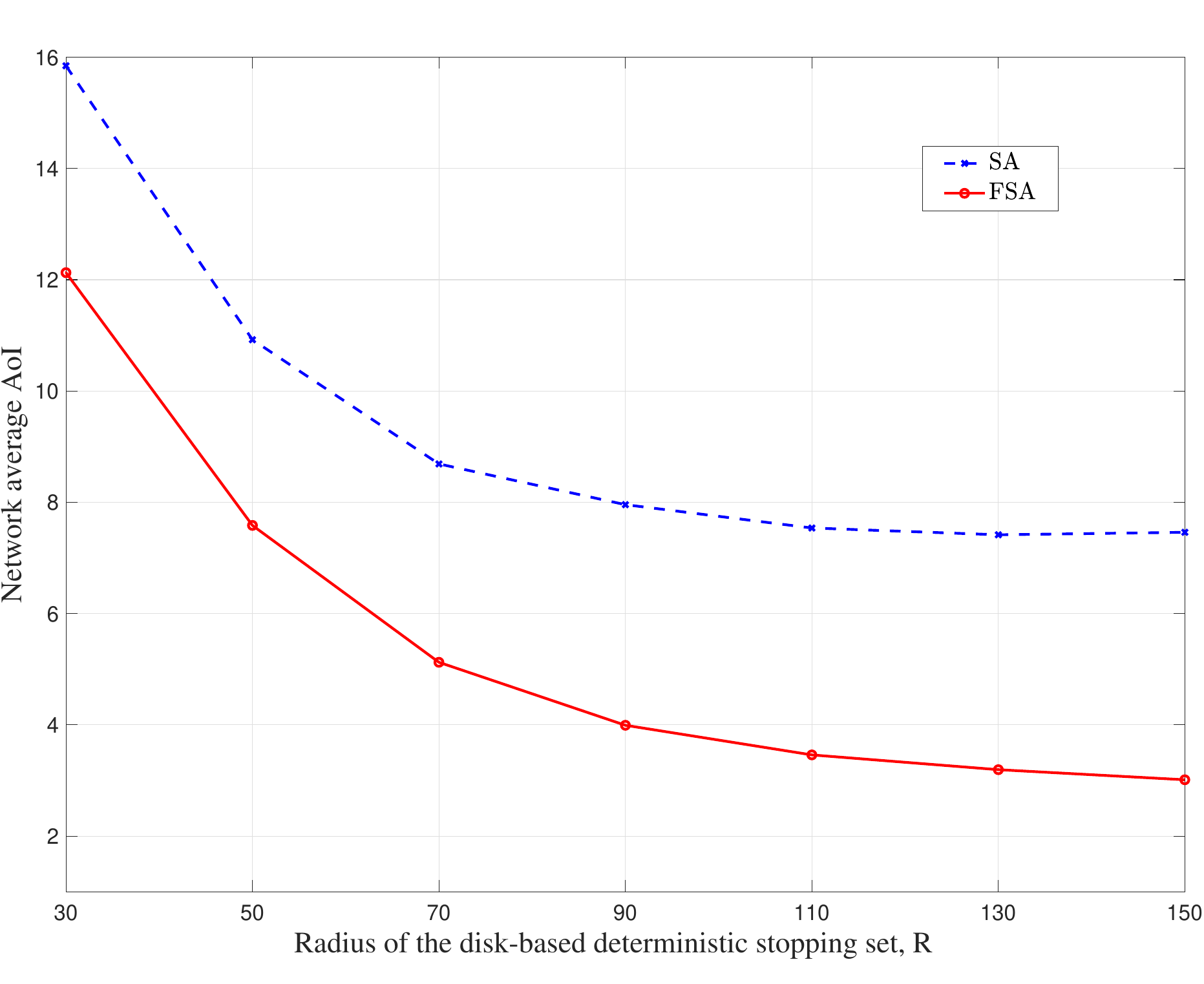}}

  \caption{Network average AoI verse radius of the deterministic stopping set, under different channel access scheme, in which we set $\lambda=3 \times 10^{-3}$.}
  \label{fig:aveAoI_R}
\end{figure}

Figure~\ref{fig:aveAoI_R} compares the network average AoI obtained under the locally adaptive FSA policy and the locally adaptive SA policy proposed in \cite{2022Locally}, for a deterministic stopping set that is a disk of radius $R$. 
The figure shows that the network average AoI decreases monotonically with the radius $R$.
This can be attributed to the fact that as $R$ increases, the nodes' observation regions expand, enabling them to obtain more information about the geographically proximate links to adjust the updating policy. In addition, we notice that the FSA protocol attains a significant reduction in terms of AoI compared to SA. 
This is because when the update rate is configured to be the same on average per time slot, a more regular update pattern can improve the performance of AoI.

\begin{figure}[t!] 
  \centering{}

    {\includegraphics[width=0.9\columnwidth]{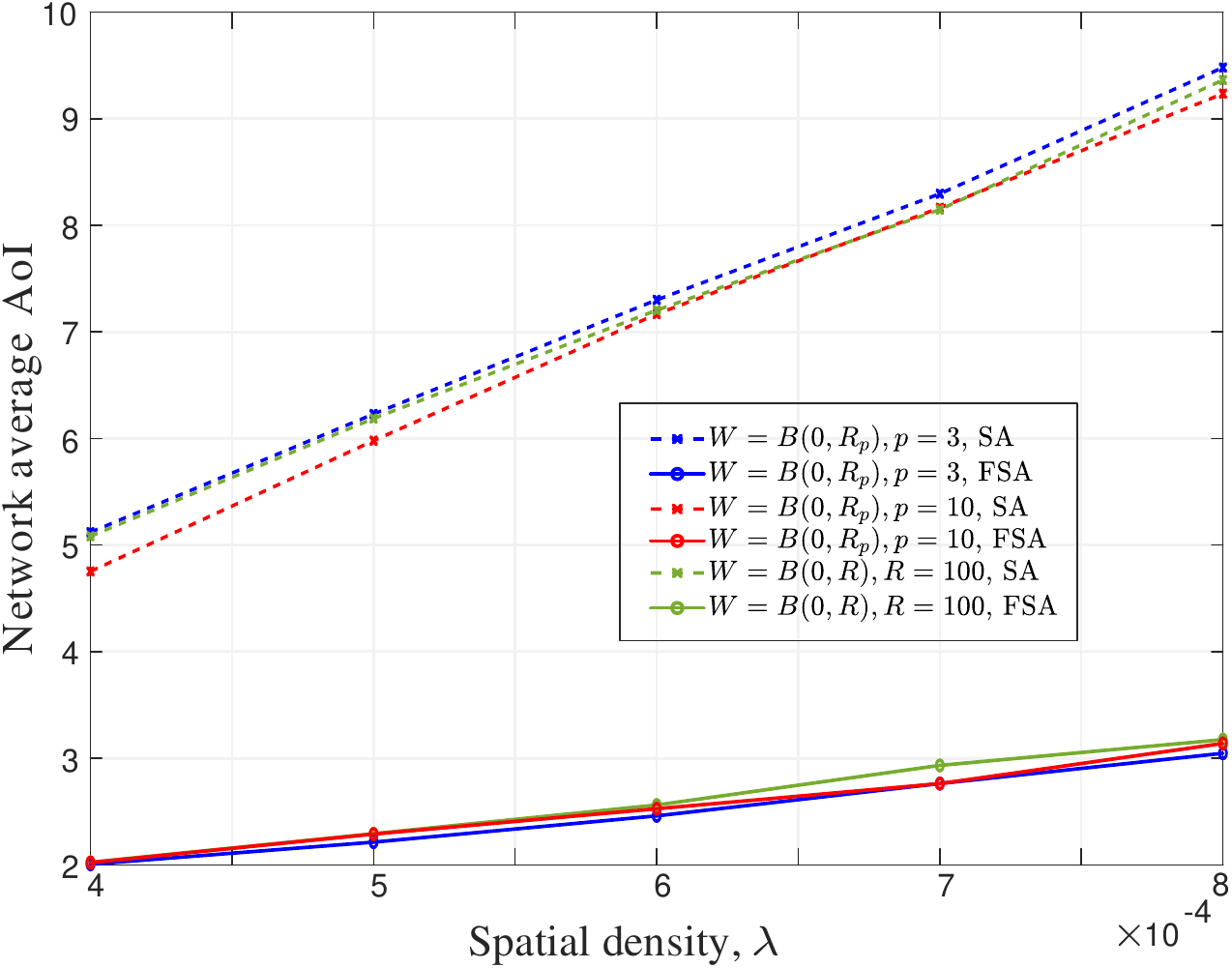}}

  \caption{Network average AoI versus spatial density under different status update policies and different stopping set $W$, in which we vary the number of geographically proximate nodes in random stopping sets as $p=3$ and $10$, and set the deterministic stopping set radius as $R=400$.}
  \label{fig:aveAoI}
\end{figure}

Figure~\ref{fig:aveAoI} plots the network average AoI as a function of the spatial deployment density under locally adaptive SA and FSA policies, with different levels of local observations (reflected by the capacity of the stopping sets).
Particularly, we consider two types of stopping sets: the random stopping set (i.e. it is a disk with a variable radius $R_p$, containing $p$ closest neighbors of a generic source node) and deterministic stopping set (i.e. it is a disk of fixed radius $R$). 
From this figure, we observe that despite an increase in spatial density inevitably deteriorating the AoI, as spatial contentions for spectral resources become more intense, the FSA protocol can reduce AoI by more than half compared to the SA-based method, whereas the gain becomes more pronounced in the very dense deployment regime (i.e. $\lambda = 1.1 \times 10^{-3}$).
In addition, when the number of observation neighbors increases (e.g., from $3$ to $10$), the network AoI performance will also be improved.

% From this figure, we can observe that an increase in node density always raises the network average AoI, but our policy is able to mitigate this deterioration.

\begin{figure}[t!] 
  \centering{}

    {\includegraphics[width=0.9\columnwidth]{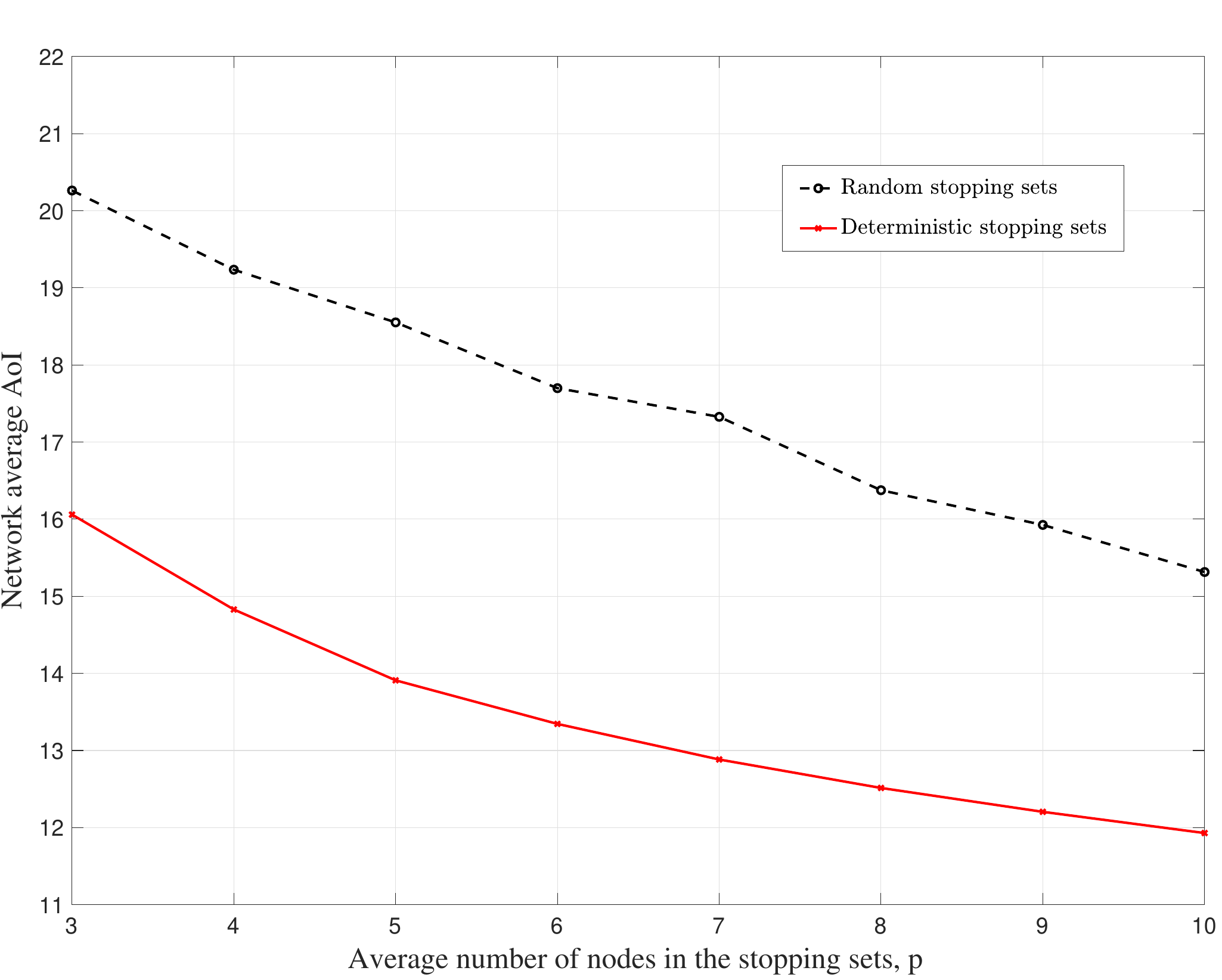}}

  \caption{Network average AoI versus the average number of nodes in both random and deterministic stopping sets under locally adaptive FSA protocol, in which we set $\lambda=5 \times 10^{-3}$.}
  \label{fig:aveAoI_p}
\end{figure}

Finally, we draw Figure~\ref{fig:aveAoI_p} to comprehend the performance of the locally adaptive FSA policy under the two types of stopping sets by comparing their network AoI performance with the same average number of sensed nodes.
Specifically, we increase the number of observed nodes $p$ in the random stopping sets from $3$ to $10$; we also calculate the corresponding detection window radius $R$ (approximately from $34.55$ to $63.08$) under the deterministic stopping sets as $R = \sqrt{\frac{p}{\lambda \pi}}$.
From this figure, we can see that when the knowledge of the surrounding environment increases, the network AoI drops significantly in both window modes.
Moreover, with the same average number of observed nodes, the deterministic stopping sets outperform the random ones.
This is because in the densely distributed region, sources need to observe more information for designing a good update strategy, whilst in the sparse region, such requirements are not so stringent. 
% the demands are less.
The deterministic stopping sets have the natural property of satisfying this spatial variation requirement, which can alleviate the interference coupling effect caused by the uneven distribution of nodes in large-scale networks.

%========================================%
%                Conclusion
%========================================%
\section{Conclusion} \label{Sec:Conclusion}
We proposed an FSA-based status update policy to minimize network average AoI with source nodes' local observations.
The policy has low complexity and can be implemented in a distributed manner. 
Specifically, it dynamically adjusts the frame size of each node according to its ambient transmission condition, effectively mitigating interference among geographically proximate links and thereby improving AoI performance. 
% thereby reducing interference amongst geographically close links and improving AoI performance. 
Additionally, we derived an analytical expression for the network average AoI under this scheme. 
We conducted numerical simulations to examine the performance of the proposed policy. 
We showed that the locally adaptive FSA policy significantly reduces the network average AoI compared to the SA counterpart. Moreover, increasing the information about the ambient environment obtained by sources will further enhance AoI performance.

Although the policy developed in this paper effectively reduces the network average AoI, it is obtained as a suboptimal solution to the targeted optimization problem. Investigating efficient algorithms to solve for the exact optimal solution can be a concrete extension of this work.

% We carried out an analytical study toward designing the status updating policy under FSA-like protocol on AoI performance in Poisson networks.
% Our policy empowered sources with local observation and allowed them to formulate the appropriate framesizes and updating rates according to their ambient environment, thus reducing interference amongst geographically close links. 
% We analyzed the derivation of the optimal updating policy and offered a suboptimal solution. 
% Under this solution, we analyzed the network average AoI. 
% Simulations were undertook on the derived upper bound and revealed that employing locally adaptive FSA yields better performance than corresponding SA.
% In addition, more information about the ambient environment gained by sources will further enhance AoI performance.

%that although the suboptimal solution does not capture the role of local information under FSA, the results still show that employing FSA yields better performance than SA. 
%In future work, we can further think about solving the optimal updating strategy to explore the effect of local observation and achieve the minimization of AoI.

%========================================%
%            Appendix
%========================================%
\begin{appendix}
\subsection{Proof of Lemma 3} \label{pro:lem3}
Deconditioning $\mu_0^\Phi$ in \eqref{equ:con_AoI}, we get the average AoI as
\begin{align} \label{equ:aveAoI}
    \bar{\Delta}_0=\frac{F_0^2-1}{12F_0}\eta_0 \mathbb{E}[\mu_0^\Phi\vert W] + \frac{F_0}{\eta_0} \mathbb{E}\bigg[\frac{1}{\mu_0^\Phi}\vert W\bigg]+\frac{1-F_0}{2}.
\end{align}
Given a stopping set $W$, by leveraging Lemma~\ref{lem:mu_0}, we calculate the expectation of $\mu_0^\Phi$ as
\begin{align} \label{equ:bipo_a1}
   &\mathbb{E}[\mu_0^\Phi \vert W]
   \nonumber\\
   &= e^{-\! \frac{\theta r^\alpha}{\rho}}\!\!\! \prod_{\substack{j \neq 0,\\ X_j \in W}}\!\! \left(1\!-\!\frac{\eta_j/F_j}{1 \!+\! D_{j0}} \right) \!\times\! \mathbb{E} \bigg[\!\! \prod_{X_j \in \mathbb{R}^2 \backslash W}\!\! \bigg(1\!-\!\frac{\eta_j/F_j}{1\!+\! D_{j0}} \bigg)\! \bigg].
\end{align}
By applying the mass transportation theorem of stationary point process \cite{BacBla:09}, we rewrite this expression as follows
\begin{align} \label{equ:bipo_a2}
   &\mathbb{E}[\mu_0^\Phi \vert W]
   \nonumber\\
   &= e^{-\! \frac{\theta r^\alpha}{\rho}}\!\!\! \prod_{\substack{j \neq 0,\\ y_j \in W}}\!\! \left(1\!-\!\frac{\eta_0/F_0}{1 \!+\! D_{0j}} \right) \!\times\! \mathbb{E} \bigg[\!\! \prod_{y_j \in \mathbb{R}^2 \backslash W}\!\! \bigg(1\!-\!\frac{\eta_0/F_0}{1\!+\! D_{0j}} \bigg)\! \bigg]
   \nonumber\\
   &= e^{-\! \frac{\theta r^\alpha}{\rho}}\!\!\! \prod_{\substack{j \neq 0,\\ y_j \in W}}\!\! \left(1\!-\!\frac{\eta_0/F_0}{1 \!+\! D_{0j}} \right)
   \!\times\! \mathbb{E} \bigg[ e^{\sum_{y_j \in \mathbb{R}^2 \backslash W}\! \log \big(1-\frac{\eta_0/F_0}{1+ D_{0j}} \big)}  \bigg].
\end{align}
According to the probability generating functional (PGFL) \cite{2022Spatio}, we know that for PPP, the following equation holds
\begin{align} \label{equ:PGFL}
    \mathbb{E} \Big[\exp \big(-s \sum_{x \in \Phi}f(x) \big) \Big] = \exp \Big(- \int_{x \in \Phi}\big[1-e^{-f(x)}\big] \Lambda \mathrm{d}x \Big),
\end{align}
where $\Lambda$ is the intensity of the PPP. By applying this, we can perform the following derivation
\begin{align} \label{equ:bipo_a3}
    &\mathbb{E}[\mu_0^\Phi \vert W]  
    % \nonumber\\
    % &=e^{-\! \frac{\theta r^\alpha}{\rho}}\!\!\! \prod_{\substack{j \neq 0,\\ X_j \in W}}\!\! \left(1\!-\!\frac{\eta_j/F_j}{1 \!+\! D_{j0}} \right) \!\times\! \mathbb{E} \bigg[\!\! \prod_{X_j \in \mathbb{R}^2 \backslash W}\!\! \bigg(1\!-\!\frac{\eta_j/F_j}{1\!+\! D_{j0}} \bigg)\! \bigg]
    % \nonumber\\
    % &\stackrel{(a)}{=} e^{-\! \frac{\theta r^\alpha}{\rho}}\!\!\! \prod_{\substack{j \neq 0,\\ y_j \in W}}\!\! \left(1\!-\!\frac{\eta_0/F_0}{1 \!+\! D_{0j}} \right) \!\times\! \mathbb{E} \bigg[\!\! \prod_{y_j \in \mathbb{R}^2 \backslash W}\!\! \bigg(1\!-\!\frac{\eta_0/F_0}{1\!+\! D_{0j}} \bigg)\! \bigg]
    % \nonumber\\
    % &\qquad \qquad \times \exp \!\bigg(\!\!-\!\!\lambda \!\int_{x \in \mathbb{R}^2 \backslash W}\!\! \bigg[1- e^{\log \Big(1\!-\!\frac{\eta_0/F_0}{1 \!+\! D_{0j}}\Big) \bigg] \!\bigg)
    \nonumber\\
    &= e^{-\! \frac{\theta r^\alpha}{\rho}}\!\!\!\! \prod_{j \neq 0, y_j \in W}\!\!\! \left(1\!-\!\frac{\eta_0/F_0}{1+D_{0j}} \right)
    \nonumber\\
    &\qquad \quad \times \exp \!\bigg(\!\!-\! \lambda \!\int_{x \in \mathbb{R}^2 \backslash W}\!\! \bigg[1- \Big(1\!-\!\frac{\eta_0/F_0}{1+\Vert x \Vert ^\alpha/\theta r^\alpha}\Big) \bigg]\mathrm{d}x \!\bigg)
    \nonumber\\
    &=e^{-\! \frac{\theta r^\alpha}{\rho}} \mathbb{E}\Bigg[ \exp \bigg(\!\!-\!\lambda \int_{x\in \mathbb{R}^2} \frac{\eta_0 /F_0 \!\!\quad \mathrm{d}x }{1+\Vert x \Vert ^\alpha/\theta r^\alpha} \bigg)\vert W\Bigg]
    \nonumber\\
    &\stackrel{(a)}{=} e^{-\! \frac{\theta r^\alpha}{\rho}} \mathbb{E}\Bigg[\exp \Big(- \lambda \pi \delta \int_0^{\infty} \frac{\eta_0/F_0}{1+u/\theta r^\alpha} u^{\delta-1} \mathrm{d}u \Big)\vert W\Bigg]
    \nonumber\\
    &\stackrel{(c)}{=} \mathbb{E}\Bigg[\exp \Big(\!\!-\!\! \frac{\theta r^\alpha}{\rho} \!\!-\!\! \lambda \pi r^2 \theta^\delta \Gamma(1\!-\!\delta) \Gamma(1\!+\!\delta) \frac{\eta_0}{F_0} \Big)\vert W\Bigg],
\end{align}
where (a) changes variables from rectangular to polar coordinates and sets $u= \Vert x \Vert ^\alpha$, 
(b) is due to the result $\int_0^{\infty} \frac{u^{\delta-1}\mathrm{d}u}{m+u}=m^{\delta-1}\frac{\pi}{\sin{(\pi \delta)}}$ \cite{2015The}.
Similarly, we have
\begin{align} \label{equ:bipo_2}
    &\mathbb{E}\bigg[\frac{1}{\mu_0^\Phi}\vert W\bigg] 
    % =e^{-\! \frac{\theta r^\alpha}{\rho}} \mathbb{E} \Bigg[\prod_{i \neq 0}\left(  1-\frac{\eta_i /F_i}{1+\Vert X_i \Vert ^{\alpha}/\theta r^\alpha} \right)^{-1} \vert W\Bigg]
    \nonumber\\
    &=e^{\! \frac{\theta r^\alpha}{\rho}}\!\!\! \prod_{\substack{j \neq 0,\\ X_j \in W}}\!\! \left(1\!-\!\frac{\eta_j/F_j}{1 \!+\! D_{j0}} \right)^{-1} \!\!\!\times\! \mathbb{E} \bigg[\!\! \prod_{X_j \in \mathbb{R}^2 \backslash W}\!\! \bigg(1\!-\!\frac{\eta_j/F_j}{1\!+\! D_{j0}} \bigg)^{-1} \bigg]
    \nonumber\\
    &=e^{\! \frac{\theta r^\alpha}{\rho}}\!\!\! \prod_{\substack{j \neq 0,\\ y_j \in W}}\!\! \left(1\!-\!\frac{\eta_0/F_0}{1 \!+\! D_{0j}} \right)^{-1} \!\!\!\times\! \mathbb{E} \bigg[\!\! \prod_{y_j \in \mathbb{R}^2 \backslash W}\!\! \bigg(1\!-\!\frac{\eta_0/F_0}{1\!+\! D_{0j}} \bigg)^{-1} \bigg]
    \nonumber\\
    &=\frac{ \exp \bigg( {\! \frac{\theta r^\alpha}{\rho} + \lambda \!\int_{x \in \mathbb{R}^2 \backslash W}\!\! \frac{\eta_0/F_0 \quad \!\! \mathrm{d}x}{1-\frac{\eta_0}{F_0} + \!\Vert x \Vert ^\alpha/\theta r^\alpha}} \bigg) }{\prod_{\substack{j \neq 0,\\ y_j \in W}}\!\! \left(1\!-\!\frac{\eta_0/F_0}{1 \!+\! D_{0j}} \right)}
    % \nonumber\\
    % &= \mathbb{E} \Bigg[\! \exp \bigg(\!\frac{\theta r^\alpha}{\rho}\!\!-\lambda \!\!\int_{x\in \mathbb{R}^2}\!\! \left[ 1\!\!-\!\!\left( 1 \!\!-\!\! \frac{\eta_0 /F_0}{1\!+\!\Vert x \Vert ^\alpha/\theta r^\alpha} \right)^{\!\!-1} \right] \mathrm{d}x \bigg) \vert W \Bigg]
    % \nonumber\\
    % &= \mathbb{E} \Bigg[\! \exp \bigg(\!\frac{\theta r^\alpha}{\rho}\!\!+\lambda \!\!\int_{x\in \mathbb{R}^2} \frac{\eta_0/F_0}{1-\frac{\eta_0}{F_0} + \frac{\Vert x \Vert ^\alpha}{\theta r^\alpha}} \mathrm{d}x \bigg) \vert W \Bigg]
    \nonumber\\
    &= \mathbb{E} \Bigg[\! \exp \bigg(\!\frac{\theta r^\alpha}{\rho}\!+\! \lambda \pi \delta \int_0^{\infty} \frac{\eta_0/F_0}{1-\frac{\eta_0}{F_0}+\frac{u}{\theta r^\alpha}} u^{\delta-1} \mathrm{d}u \bigg) \vert W \Bigg]
    \nonumber\\
    &= \mathbb{E} \Bigg[\! \exp \bigg(\!\frac{\theta r^\alpha}{\rho}\!\!+\!\! \lambda \pi r^2 \theta^\delta \Gamma(1\!\!-\!\!\delta) \Gamma(1\!\!+\!\!\delta) \frac{\eta_0}{F_0} \Big(1\!\!-\!\!\frac{\eta_0}{F_0} \Big)^{\delta-1} \bigg) \vert W \Bigg].
\end{align}
Substituting \eqref{equ:bipo_a3} and \eqref{equ:bipo_2} into \eqref{equ:aveAoI}, we can obtain the result in Lemma 3.
\end{appendix}

%\balance
\bibliographystyle{IEEEtran}
\bibliography{bib/StringDefinitions,bib/IEEEabrv,bib/howard_AoI_Ctrl}

\end{document}